\documentclass[journal,twoside,web]{ieeecolor}
\usepackage{generic}
\usepackage{cite}
\usepackage{amsmath,amssymb,amsfonts}
\usepackage{algorithmic}
\usepackage{graphicx}
\usepackage{algorithm,algorithmic}
\usepackage{hyperref}
\usepackage{textcomp}
\def\BibTeX{{\rm B\kern-.05em{\sc i\kern-.025em b}\kern-.08em
    T\kern-.1667em\lower.7ex\hbox{E}\kern-.125emX}}
\usepackage{multirow}
\usepackage{array}
\usepackage{float}
\usepackage{url}
\usepackage{makecell}
\usepackage{MnSymbol}
\newcommand{\squotes}[1]{`#1'}

\newcommand{\etal}{et al.\ }

\newcommand{\fref}[1]{Fig.~\ref{#1}}
\newcommand{\tref}[1]{Table~\ref{#1}}
\newcolumntype{$}{>{\global\let\currentrowstyle\relax}}
\newcolumntype{^}{>{\currentrowstyle}}
\newcommand{\setrow}[1]{\gdef\currentrowstyle{#1}%
  #1\ignorespaces
}
\begin{document}
\title{Single-word Auditory Attention Decoding Using Deep Learning Model}

\author{Nhan Duc Thanh Nguyen$^{1}$, Huy Phan$^{2}$, Kaare Mikkelsen$^{1}$ and Preben Kidmose$^{1}$
\thanks{$^{1}$Nhan Duc Thanh Nguyen (corresponding author, e-mail: {\tt\small ndtn@ece.au.dk}), Kaare Mikkelsen (e-mail: {\tt\small mikkelsen.kaare@ece.au.dk}) and Preben Kidmose (e-mail: {\tt\small pki@ece.au.dk}) are with Department of Electrical and Computer Engineering, Aarhus University, 8200 Aarhus N, Denmark.}
\thanks{$^{2}$Huy Phan is with Amazon, Cambridge, MA 02142, USA
        (e-mail: {\tt\small huypq@amazon.com}). The work does not relate to H.P.'s work at Amazon.}%
}

\maketitle

\begin{abstract}
Identifying auditory attention by comparing auditory stimuli and corresponding brain responses, is known as auditory attention decoding (AAD). The majority of AAD algorithms utilize the so-called envelope entrainment mechanism, whereby auditory attention is identified by how the envelope of the auditory stream drives variation in the electroencephalography (EEG) signal. However, neural processing can also be decoded based on endogenous cognitive responses, in this case, neural responses evoked by attention to specific words in a speech stream. This approach is largely unexplored in the field of AAD but leads to a single-word auditory attention decoding problem in which an epoch of an EEG signal timed to a specific word is labeled as attended or unattended. This paper presents a deep learning approach, based on EEGNet, to address this challenge. We conducted a subject-independent evaluation on an event-based AAD dataset with three different paradigms: \textit{word category oddball}, \textit{word category with competing speakers}, and \textit{competing speech streams with targets}. The results demonstrate that the adapted model is capable of exploiting cognitive-related spatiotemporal EEG features and achieving at least 58\% accuracy on the most realistic competing paradigm for the unseen subjects. To our knowledge, this is the first study dealing with this problem.
\end{abstract}


\section{Introduction}
\IEEEPARstart{T}{he} human brain possesses a remarkable ability to perceptually segregate concurrent auditory objects and selectively attend to the objects of interest. Identifying these objects, based on electrophysiological signals such as electroencephalography (EEG), electrocorticography (ECoG), and magnetoencephalography (MEG), is commonly referred to as auditory attention decoding (AAD). Current state-of-the-art AAD studies have focused on an audio envelope reconstruction approach in which the neural response to auditory attention is considered as entrainment to the audio envelope of the attended auditory stream \cite{osullivanAttentionalSelectionCocktail2015a, biesmansAuditoryInspiredSpeechEnvelope2017a, detaillezMachineLearningDecoding2020a, ciccarelliComparisonTwoTalkerAttention2019a, thorntonRobustDecodingSpeech2022}. The attended stream is identified as the stream whose envelope exhibits the highest correlation with the reconstructed envelope. These methods require detailed knowledge of stimuli and are primarily based on the exogenous response. Recently, a new study, which focuses on the cognitive responses to specific auditory events in a stream \cite{nguyenStudyCognitiveComponent2023}, has shown that the cognitive processing of single-word speech events in a multi-talker environment elicits an event-related potential (ERP). This finding opens the door to a simplified version of the AAD problem, involving classifying a single EEG epoch around a word onset as either attended or unattended. In this paper, we investigate the feasibility of this approach.

As mentioned earlier, this is a different approach to address the AAD problem compared to the envelope following response \cite{osullivanAttentionalSelectionCocktail2015a, biesmansAuditoryInspiredSpeechEnvelope2017a, detaillezMachineLearningDecoding2020a, ciccarelliComparisonTwoTalkerAttention2019a, thorntonRobustDecodingSpeech2022}. However, the proposed approach can be compared to single-trial ERP classification, as addressed in various Brain-Computer Interface (BCI) applications. One of the very first published BCI systems, based on a cognitive evoked potential known as the P300 component, was the so-called \textit{P300 speller}, introduced by Farwell \etal in 1988 \cite{farwellTalkingTopYour1988}. Since then, various methods have been proposed using different signal processing algorithms and conventional classifiers such as an ensemble of support vector machines \cite{rakotomamonjyBCICompetitionIII2008}, gradient-boosting \cite{hoffmannBoostingApproachP3002005}, linear discriminant analysis\cite{blankertzOptimizingSpatialFilters2008, hoffmannEfficientP300basedBraincomputer2008a, blankertzSingletrialAnalysisClassification2011}, ICA \cite{xuBCICompetition2003data2004, serbyImprovedP300basedBraincomputer2005}, and xDAWN \cite{rivet*XDAWNAlgorithmEnhance2009}. Recently, the development of algorithms for single-trial ERP classification has been significantly influenced by deep learning methods. The convolutional neural networks (CNNs) approach has been reported to outperform the other methods when the number of electrodes is restricted to 8 \cite{cecottiConvolutionalNeuralNetworks2011}. In 2018, Lawhern \etal introduced EEGNet, a compact convolutional neural network \cite{lawhernEEGNetCompactConvolutional2018}, which can extract EEG features and generalize across BCI paradigms. Shi \etal \cite{shiConvolutionalLSTMNetwork2015} proposed a convolutional long short-term memory (ConvLSTM) architecture that could effectively capture spatiotemporal correlations for predicting rainfall. This architecture was later integrated into ensemble models \cite{joshiSingleTrialP3002018} to achieve better performance in P300 speller classification. However, compared to the EEGNet, the drawback of ConvLSTM, and other modern structures such as the transformer, is that it, in general, requires more data to train. Several approaches have attempted to improve the performance by combining recurrent neural network (RNN) with CNN such as CNN-RNN-Net \cite{tuleuovDeepLearningModels2019} and CNN-LSTM \cite{leoniSingletrialStimuliClassification2022}. However, while these models have tended to be larger than EEGNet, they have not exhibited superiority over the EEGNet in generalization.

In this study, we investigate the feasibility of single-word attention classification, based on the data set described in \cite{nguyenStudyCognitiveComponent2023}. However, due to its specialized nature, the data set is quite small ($< 6000$ attended data points across all three paradigms) and imbalanced ($\approx$ 1:5 ratio between the two classes), which motivates several of the design decisions in our approach:
\begin{itemize}
    \item We focus on the EEGNet architecture, due to its high performance and relatively small size.
    \item We propose two data augmentation techniques that turn out to be important for training the model.
\end{itemize}
As will be seen below, we find that the problem of single-word attention classification is challenging, but not impossible. We compare our proposed approach to the conventional linear model of the envelope-based approach \cite{osullivanAttentionalSelectionCocktail2015a} using the same analysis window.

\section{Materials and Methods}
\subsection{Experimental dataset}
In this work, we use data from the first three out of the four paradigms in the event-based AAD dataset \cite{nguyenStudyCognitiveComponent2023} which focuses on AAD based on the cognitive processing of speech events. Paradigm 1, \textit{word category oddball}, is similar to a conventional oddball paradigm in which the standard events and target events are short spoken words: cardinal numbers and animal names/color names, respectively. In Paradigm 2, \textit{word category with competing speakers}, similar stimuli were simultaneously presented from two loudspeakers. The participant was instructed to pay attention to only the target events in one of the streams, completely ignore the other and passively count the number of target events in the attended stream. In Paradigm 3, \textit{competing speech streams with targets}, a similar setup of two competing streams was used as in Paradigm 2. However, in each stream, audio snippets of different stories were used instead of discrete spoken words. Within each story, a category of words was designated as target words, chosen from one of four categories: \textit{animal names}, \textit{human names}, \textit{color names}, and \textit{plant species}.

EEG data from 24 normal-hearing subjects were recorded and sampled at 1000 Hz using 32 scalp electrodes. The data was pre-processed by: down-sampling to 256 Hz, applying a zero-phase FIR band-pass filter from 0.5 to 40 Hz, applying the Independent Component Analysis (ICA) method to reject eye artifacts, and epoching data from -200 milliseconds to 1 second relative to the middle of words with 200-$\mu$V peak-to-peak rejection. Finally, epochs were labeled as attended or unattended according to the experimental conditions. It is important to note that attended epochs in Paradigm 2 and 3 are based on the target events in the attended stream and unattended epochs are based on all the events in the unattended stream. In the end, for each subject, the number of attended and unattended epochs across the three paradigms (before applying epoch rejection) are 60 and 260, 74 and 400, 109 and 1172, respectively. More details on the experimental protocol and setup can be found in the original study \cite{nguyenStudyCognitiveComponent2023}. Hereafter, this dataset is referred to as EventAAD data.


\subsection{Data augmentation}\label{ss_data_augmentation}
As mentioned above, the EventAAD dataset is relatively small and inherently imbalanced due to the oddball design. This is a common challenge in BCI systems data, and the challenge with limited data is even more pressing with a deep learning model as the number of parameters increases. We address these problems with two data augmentation methods tailored to ERPs: upsampling by averaging and simulation of target ERPs.

\textit{Upsampling by averaging} method is inspired by the traditional averaging method of ERP quantification. For each subject and for each class (attended / unattended), a new epoch is created by averaging $k$ random epochs drawn without replacement from that class. In this work, $k$ was randomly chosen from 1 to 3. It is important to note the effect of $k$: as it increases, a clearer ERP pattern is induced into the training data. Besides giving a greater variation in the signal-to-noise ratios seen by the model, it potentially helps the model focus on the most promising patterns in the data. In the case of $k = 1$, the randomly drawn epochs are merely duplicated. Upsampling was repeated until the number of epochs in each class was double the number of unattended epochs in the original dataset, resulting in a new dataset called the \textit{upsampled experimental dataset}.

\begin{figure*}[t]
\vspace{-10pt}
\centering
\includegraphics[width=1.0\linewidth]{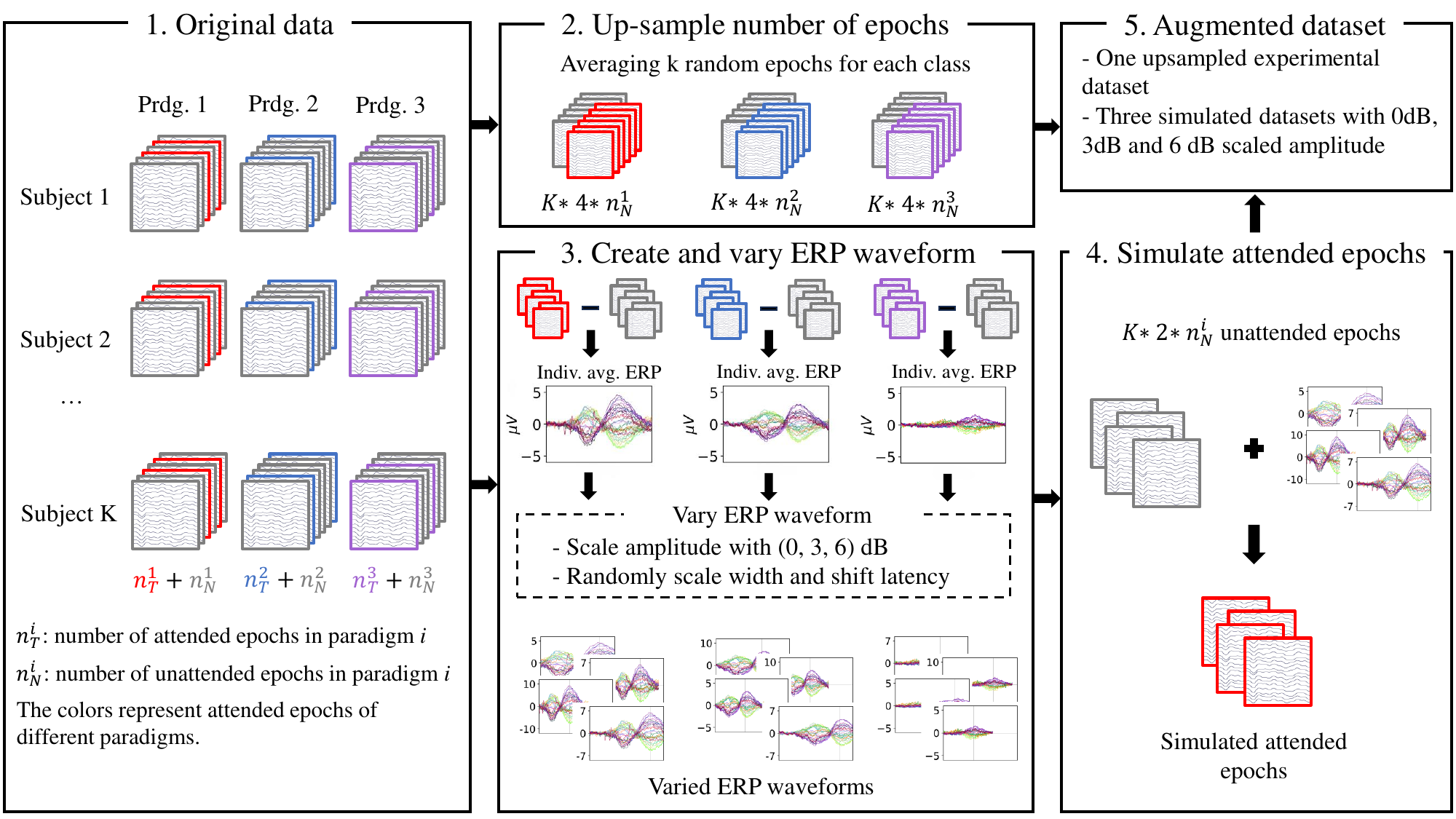}
\caption{Visual overview of the data augmentation methods. \textbf{1} The original data after pre-processing and epoching. \textbf{2} Data of each paradigm is up-sampled by averaging 1 to 3 (duplicating when $k = 1$) random epochs within a class to create an upsampled experimental dataset. \textbf{3} Compute an individual average difference ERP waveform for each paradigm and vary the ERP waveforms by scaling amplitude, width, and shifting latency. \textbf{4} Simulate new attended epochs for each paradigm by adding each unattended epoch and a varied ERP waveform. \textbf{5} Create an augmented dataset including one upsampled experimental dataset and three simulated datasets.}
\label{fig1_data_augmentation}
\vspace{-10pt}
\end{figure*}

\textit{ERP simulation} method was introduced by Depuydt \etal \cite{depuydtSingletrialERPQuantification2023} for training neural networks to quantify ERPs. The method is to synthesize a new target (attended) epoch by adding a target (attended) ERP waveform at a known latency to a random non-target (unattended) epoch. In this work, instead of using a half-cycle sinusoidal wave as in the original study, the attended ERP waveform was estimated individually by averaging all attended epochs for each subject to closely resemble the experimental data. To take the inter-trial variability into account, the latency of the ERP waveform was uniformly sampled around the real latency of the individual ERP with a standard deviation of 10 ms. The width of the ERP waveform was uniformly sampled between 300 ms and 600 ms, which is the expected range for the P300 component in literature \cite{polichUpdatingP300Integrative2007, nguyenStudyCognitiveComponent2023}. Variation of the ERP amplitude was induced by randomly amplifying the ERP amplitude by 0 dB, 3 dB, and 6 dB. To increase the number of epochs in the simulated dataset, first, the unattended epochs were upsampled by a factor of 4. Subsequently, half of these scaled epochs were used to synthesize new attended epochs. A visual overview of the augmentation process is shown in \fref{fig1_data_augmentation}.

In summary, two augmentation methods were applied to the data of three paradigms for each subject to create an augmented dataset which included one upsampled experimental (upsampled exp.) dataset and three simulated datasets corresponding to 3 SNR levels: 0 dB, 3 dB, and 6 dB. Details of these datasets are shown in \tref{tabl1_dataset}

\begin{table}[h]
\caption{\label{tabl1_dataset}Details of the number of data points in original and augmented datasets.}
\vspace{-10pt}
\begin{center}
\resizebox{\columnwidth}{!}{\begin{tabular}{$c|^c^c^c^c^c^c} 
 \hline
 \setrow{\bfseries}\multirow{2}{*}{Dataset} & \multicolumn{2}{^c}{Paradigm 1} & \multicolumn{2}{^c}{Paradigm 2} & \multicolumn{2}{^c}{Paradigm 3}\\
 \setrow{\bfseries}\multirow{2}{*} & attd. & unattd. & attd. & unattd. & attd. & unattd.\\ 
 \hline
 Original & 1440 & 6240 & 1776 & 9600 & 2611 & 28128 \\
 
 Upsampled exp. & 12480 & 12480 & 19200 & 19200 & 56256 & 56256 \\ 
 
 Simulated 0dB & 12480 & 12480 & 19200 & 19200 & 56256 & 56256 \\ 

 Simulated 3dB & 12480 & 12480 & 19200 & 19200 & 56256 & 56256 \\ 

 Simulated 6dB & 12480 & 12480 & 19200 & 19200 & 56256 & 56256 \\ 

 \setrow{\bfseries}Augmented & 49920 & 49920 & 76800 & 76800 & 225024 & 225024 \\ 
 \hline
\end{tabular}}
\end{center}
\vspace{-10pt}
\end{table}

\subsection{EEGNet}
The proposed network is a compact CNN architecture with depthwise and separable convolutions, adapted from the EEGNet \cite{lawhernEEGNetCompactConvolutional2018}, originally introduced for EEG-based BCIs. The network comprises two main convolutional blocks and one fully connected classification layer. The first block contains $F_1$ 2D frequency-oriented convolutional filters of kernel size (1, $K_1$) to capture the frequency information followed by $D \cdot F1$ depthwise spatial-oriented convolution of size ($C$, 1) to learn a spatial filter. $K_1$ is set at half the sampling rate. $D$ is a depth parameter to control the number of spatial filters. $C$ is the number of EEG channels. The next layer sequentially includes a batch normalization, an exponential linear unit (ELU) activation, an average pooling, and a dropout layer. In the second block, to decouple the relationship between feature maps outputted by the first block, it is designed to have $F_2$ separable convolutional filters of size (1, $K_2$), followed by batch normalization, ELU activation, average pooling, and dropout. Finally, the output of the second block is passed to the classification layer which consists of a fully-connected layer and a sigmoid activation. Here, to fit the model to a 32-channel, 256 Hz sampling rate dataset, the following set of parameters: $F_1 = 8$, $K_1 = 128$, $D = 2$, $C = 32$, $F_2 = 32$ and $K_2 = 16$ were used.

\subsection{Training and validation}
We performed two comparisons:
\begin{itemize}
    \item The performance of the models trained with and without data augmentation.
    \item The performance of the paradigm-independent model (trained on the combined augmented dataset) and paradigm-specific models (trained separately on the augmented data of each paradigm)
\end{itemize}
The performance of each model was evaluated by two validation schemes: multi-fold cross validation and leave-one-subject-out (LOSO) validation.

\subsubsection{Multi-fold cross validation}
An 8-fold cross-validation was performed for each model. The data was split into 8 folds. 7 out of the 8 folds were split into a training set and a validation set with a ratio of 4:1 to train the model while the remaining fold was held out for the test set. In the cases of augmented datasets, to ensure that the training set and validation set were balanced and that all unattended epochs used to simulate new attended epochs in the validation set were not seen in the training set, the 8-fold split was performed before the data augmentation and based on a trial basis across all paradigms. The trained models were then evaluated on each paradigm of the test set which was not augmented. This process was repeated for the different folds and the mean and standard deviation of accuracy were then calculated. The result of this validation is called subject-pooled performance. To compensate for the imbalance in the original test set, the accuracy for each class was first calculated. The average of these accuracies was then taken as the reported accuracy.

\subsubsection{LOSO validation}
To evaluate the performance of models on unseen subjects, a LOSO validation was performed. The data of 23 subjects was split into a training set and a validation set in a 4:1 ratio to train the model. The remaining subject was used for testing. Similar to the multi-fold cross-validation scheme, in the cases of augmented datasets, the train and validation set were augmented while the test set was kept original. This process was repeated 24 times to make sure each subject was tested once. The mean and standard deviation of accuracy across folds were then assessed. The performance of each model on each paradigm of the original data was assessed. 

The model was implemented in the PyTorch framework. The most expensive version of the model, the LOSO model for augmented data, was trained in 10 hours on a single A100 GPU. The model has roughly 2930 parameters, requiring memory of 11720 bytes to deploy. We ran 300 epochs (the number of times each data point passed through the model) and saved the models that produced the lowest binary cross-entropy loss on the validation set. The Adam optimizer, a batch size of 64, and a learning rate of $10^{-4}$ were used. Statistical tests for all comparisons were done using one-sided paired permutation tests \cite{marisNonparametricStatisticalTesting2007a}.

\section{Results}
\subsection{Subject-pooled performance}\label{ss_results_SI}
\begin{figure}[t]
\vspace{-10pt}
\centering
\includegraphics[width=1.0\linewidth]{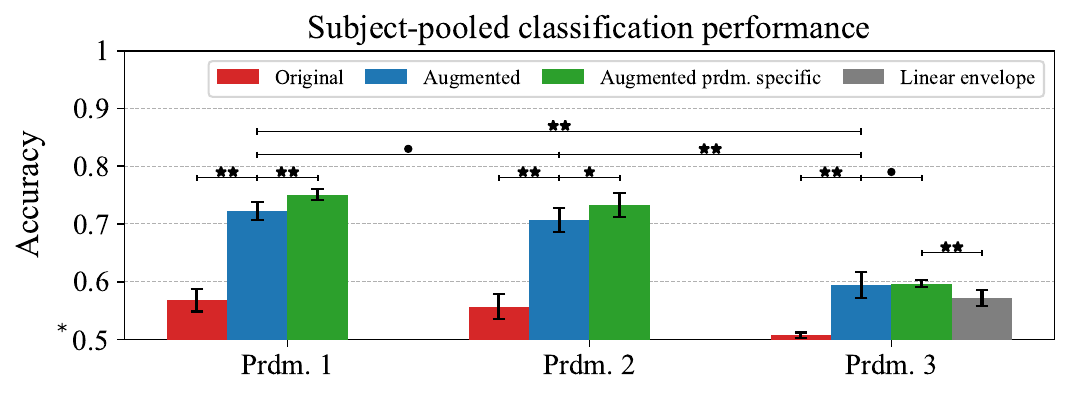}
\caption{8-fold average classification performance of models of EEGNet trained on the original dataset (paradigm-independent model without augmentation), the augmented dataset (paradigm-independent model with augmentation), the augmented data of each paradigm (paradigm-specific model with augmentation), and the envelope-based linear model (only for Paradigm 3). The error bar represents the standard deviation. \squotes{Prdm. 1}, \squotes{Prdm. 2} and \squotes{Prdm. 3} indicate the model performances on the non-augmented data of the test set of Paradigms 1, 2, and 3, respectively. The horizontal bars show the statistical test of corresponding comparisons ( $\filledstar\filledstar$: $p < 0.001$, $\filledstar$: $0.001 \le p \le 0.05$, $\bullet$: $p > 0.05$). Reported accuracy is weighted accuracy with equal weights across the two classes. (*) The accuracy of 0.5 is the chance level.}
\label{fig2_SI_performance}
\vspace{-10pt}
\end{figure}

\fref{fig2_SI_performance} shows the 8-fold average performance of the models trained on: the original dataset (paradigm-independent model without augmentation), the augmented dataset (paradigm-independent model with augmentation), the augmented data of each paradigm (paradigm-specific model with augmentation), and the envelope-based linear model (only for Paradigm 3). The columns \squotes{Prdm. 1}, \squotes{Prdm. 2} and \squotes{Prdm. 3} indicate the model performances on the non-augmented data of the test set of Paradigms 1, 2, and 3, respectively. 

Comparing models with and without data augmentation, the model trained on the original data performed significantly above the chance level but significantly worse than the model trained on the augmented dataset for all three paradigms ($p < 0.001$): 0.568 vs. 0.722 for Paradigm 1, 0.557 vs. 0.707, for Paradigm 2 and 0.507 vs. 0.594 for Paradigm 3. 

Comparing paradigm-specific and paradigm-independent models, the paradigm-specific models performed significantly better than the paradigm-independent model in Paradigm 1 (0.751 vs. 0.722, $p < 0.001$) and Paradigm 2 (0.732 vs. 0.707, $p = 0.016$) but not in Paradigm 3 (0.596 vs. 0.594, $p = 0.4$). Additionally, we compare the performance of the model (trained on the augmented dataset) on different paradigms and observe that the model performs significantly better on Paradigm 1 and 2 than on Paradigm 3 ($p<0.001$).

When comparing our model to the envelope-based linear model, we only use Paradigm 3, which is the most similar to what is done in the literature. The linear model performance is evaluated using the analysis window length of 1.2 s, corresponding to the length of the epoch in this study. We find that the paradigm-specific model significantly outperforms the linear model (0.596 vs 0.571, $p < 0.001$).

\subsection{LOSO performance}
\begin{figure}[h]
\centering
\includegraphics[width=1.0\linewidth]{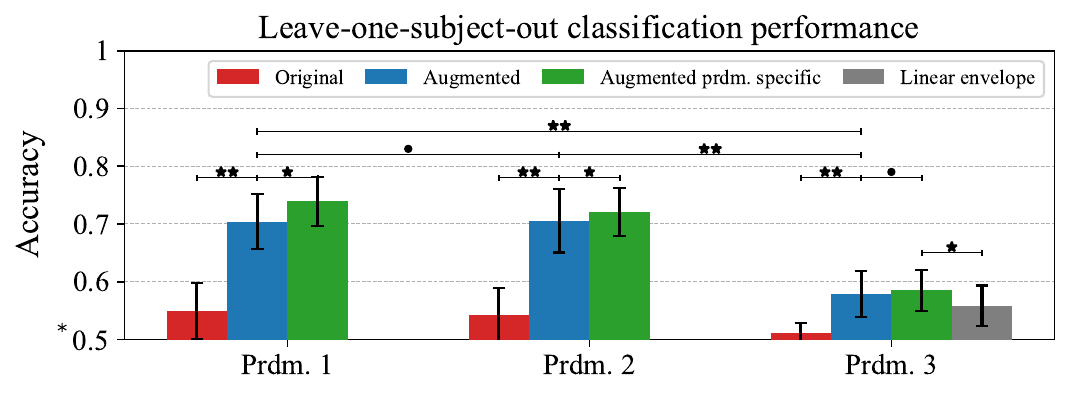}
\caption{LOSO classification performance of models of EEGNet, and the envelope-based linear model (only for Paradigm 3).}
\label{fig3_CS_performance}
\vspace{-10pt}
\end{figure}

The LOSO classification results illustrate a similar pattern as of the subject-pooled performance (see \fref{fig3_CS_performance}). The performance of the model trained on the augmented dataset is significantly higher than the one trained on the original dataset for all three paradigms ($p < 0.001$) but significantly lower than paradigm-specific models (except Paradigm 3): 0.704 vs. 0.739, $p = 0.002$ for Paradigm 1; 0.705 vs. 0.721, $p = 0.018$ for Paradigm 2 and 0.579 vs. 0.585, $p = 0.225$ for Paradigm 3. In comparison between paradigms, the model, which was trained on the augmented dataset, performed significantly better on Paradigms 1 and 2 than on Paradigm 3 ($p < 0.001$). Additionally, the paradigm-specific also outperforms the envelope linear model (0.585 vs. 0.558, $p = 0.01$)


\section{Discussion and Conclusions}
We find that it is in fact possible to decode attention on a single-word basis, however, even with a relatively small model, significant amounts of data are needed. Fortunately, the data augmentation strategy outlined here turned out to be effective. In this regard, it is interesting to note that even though the constructed epochs generally have a higher signal-to-noise ratio than the raw data, their inclusion still leads to a significant improvement in final model performance (it is crucial to remember here that data augmentation was not applied to the test data). Additionally, the \textit{upsampling by averaging} method tends to preserve the latency and width of the ERP components and is, therefore, suitable for classifying ERPs with less inter-subject variation in latency and width. The \textit{ERP simulation} method, on the other hand, introduces variability in new data points for both latency and width and could therefore be beneficial to apply for classifying highly variable ERP components across subjects.

The previous work with this data set \cite{nguyenStudyCognitiveComponent2023} found distinctly \squotes{cleaner} ERPs for Paradigms 1 and 2 compared to Paradigm 3, and, not surprisingly, we find that this makes single-word decoding easier. Additionally, the paradigm-specific models showed better performance than the paradigm-independent model, even though the three paradigms represent the same cognitive ERP component triggered by the same task as in the original work \cite{nguyenStudyCognitiveComponent2023}. A potential reason for this could be that the model’s capability to generalize across all three paradigms is limited. This suggests a direction for future work towards a more advanced model, possibly to uncover what part of the ERP the model relies on.

The LOSO validation results show a slight drop in performance for the unseen subjects compared to the performance of the subject-pooled data. This decrease in performance is likely due to the inter-subject variability in cognitive response latency and topography. Nevertheless, while this challenge is common in BCI systems, it is not insurmountable. We believe the accuracy can be improved by fine-tuning the model using individual subject data.

The results show that a relatively light-weight EEGNet-based model effectively extracts spatial and temporal features from the cognitive component triggered by single-word auditory attention. As mentioned in the introduction, the method proposed here differs from the conventional approach of detecting the envelope-following response. This difference makes it difficult to perform an apples-to-apples comparison between the results shown here and the state-of-the-art in AAD. However, we tested the performance of the linear model \cite{osullivanAttentionalSelectionCocktail2015a}, which is reported as the most consistent AAD method using the envelope feature \cite{geirnaertElectroencephalographyBasedAuditoryAttention2021}, for time windows of 1.2 seconds, and found that our proposed method obtained a consistently better performance.

It is worth noting that since this approach focuses on the endogenous response instead of the (stimuli-driven) exogenous response, a model could possibly be developed that combines this approach with the envelope-following response. We believe that this possibility is one of the main points of interest in our findings.  


\section*{Acknowledgement}
This work was funded by the William Demant Foundation, grant numbers 20-2673, and supported by Center for Ear-EEG, Department of Electrical and Computer Engineering, Aarhus University, Denmark.

\bibliography{references}

\begin{thebibliography}{10}

\bibitem{osullivanAttentionalSelectionCocktail2015a}
James~A. O'Sullivan et~al.,
\newblock ``Attentional {{Selection}} in a {{Cocktail Party Environment Can Be Decoded}} from {{Single-Trial EEG}},''
\newblock {\em Cereb Cortex}, vol. 25, no. 7, pp. 1697--1706, July 2015.

\bibitem{biesmansAuditoryInspiredSpeechEnvelope2017a}
Wouter Biesmans et~al.,
\newblock ``Auditory-{{Inspired Speech Envelope Extraction Methods}} for {{Improved EEG-Based Auditory Attention Detection}} in a {{Cocktail Party Scenario}},''
\newblock {\em IEEE Trans Neural Syst Rehabil Eng}, vol. 25, no. 5, pp. 402--412, May 2017.

\bibitem{detaillezMachineLearningDecoding2020a}
Tobias {de Taillez} et~al.,
\newblock ``Machine learning for decoding listeners' attention from electroencephalography evoked by continuous speech,''
\newblock {\em Eur J Neurosci}, vol. 51, no. 5, pp. 1234--1241, Mar. 2020.

\bibitem{ciccarelliComparisonTwoTalkerAttention2019a}
Gregory Ciccarelli et~al.,
\newblock ``Comparison of {{Two-Talker Attention Decoding}} from {{EEG}} with {{Nonlinear Neural Networks}} and {{Linear Methods}},''
\newblock {\em Sci Rep}, vol. 9, no. 1, pp. 11538, Aug. 2019.

\bibitem{thorntonRobustDecodingSpeech2022}
Mike Thornton et~al.,
\newblock ``Robust decoding of the speech envelope from {{EEG}} recordings through deep neural networks,''
\newblock {\em J. Neural Eng.}, vol. 19, no. 4, pp. 046007, July 2022.

\bibitem{nguyenStudyCognitiveComponent2023}
Nhan D.~T. Nguyen et~al.,
\newblock ``Study of cognitive component of auditory attention to natural speech events,'' Dec. 2023, arXiv:2312.10543.

\bibitem{farwellTalkingTopYour1988}
L.~A. Farwell and E.~Donchin,
\newblock ``Talking off the top of your head: Toward a mental prosthesis utilizing event-related brain potentials,''
\newblock {\em Electroencephalogr Clin Neurophysiol}, vol. 70, no. 6, pp. 510--523, Dec. 1988.

\bibitem{rakotomamonjyBCICompetitionIII2008}
Alain Rakotomamonjy and Vincent Guigue,
\newblock ``{{BCI Competition III}}: {{Dataset II- Ensemble}} of {{SVMs}} for {{BCI P300 Speller}},''
\newblock {\em IEEE. Trans. Biomed. Eng.}, vol. 55, no. 3, pp. 1147--1154, Mar. 2008.

\bibitem{hoffmannBoostingApproachP3002005}
U.~Hoffmann et~al.,
\newblock ``A {{Boosting Approach}} to {{P300 Detection}} with {{Application}} to {{Brain-Computer Interfaces}},''
\newblock in {\em Conference {{Proceedings}}. 2nd {{International IEEE EMBS Conference}} on {{Neural Engineering}}, 2005.}, Mar. 2005, pp. 97--100.

\bibitem{blankertzOptimizingSpatialFilters2008}
Benjamin Blankertz et~al.,
\newblock ``Optimizing {{Spatial}} filters for {{Robust EEG Single-Trial Analysis}},''
\newblock {\em IEEE Signal Process. Mag.}, vol. 25, no. 1, pp. 41--56, 2008.

\bibitem{hoffmannEfficientP300basedBraincomputer2008a}
Ulrich Hoffmann et~al.,
\newblock ``An efficient {{P300-based}} brain-computer interface for disabled subjects,''
\newblock {\em J Neurosci Methods}, vol. 167, no. 1, pp. 115--125, Jan. 2008.

\bibitem{blankertzSingletrialAnalysisClassification2011}
Benjamin Blankertz et~al.,
\newblock ``Single-trial analysis and classification of {{ERP}} components --- {{A}} tutorial,''
\newblock {\em NeuroImage}, vol. 56, no. 2, pp. 814--825, May 2011.

\bibitem{xuBCICompetition2003data2004}
Neng Xu et~al.,
\newblock ``{{BCI}} competition 2003-data set {{IIb}}: Enhancing {{P300}} wave detection using {{ICA-based}} subspace projections for {{BCI}} applications,''
\newblock {\em IEEE. Trans. Biomed. Eng.}, vol. 51, no. 6, pp. 1067--1072, June 2004.

\bibitem{serbyImprovedP300basedBraincomputer2005}
Hilit Serby et~al.,
\newblock ``An improved {{P300-based}} brain-computer interface,''
\newblock {\em IEEE Trans Neural Syst Rehabil Eng}, vol. 13, no. 1, pp. 89--98, Mar. 2005.

\bibitem{rivet*XDAWNAlgorithmEnhance2009}
Bertrand Rivet* et~al.,
\newblock ``{{xDAWN Algorithm}} to {{Enhance Evoked Potentials}}: {{Application}} to {{Brain}}--{{Computer Interface}},''
\newblock {\em IEEE. Trans. Biomed. Eng.}, vol. 56, no. 8, pp. 2035--2043, Aug. 2009.

\bibitem{cecottiConvolutionalNeuralNetworks2011}
Hubert Cecotti and Axel Graser,
\newblock ``Convolutional {{Neural Networks}} for {{P300 Detection}} with {{Application}} to {{Brain-Computer Interfaces}},''
\newblock {\em IEEE Trans. Pattern Anal. Mach. Intell.}, vol. 33, no. 3, pp. 433--445, Mar. 2011.

\bibitem{lawhernEEGNetCompactConvolutional2018}
Vernon~J. Lawhern et~al.,
\newblock ``{{EEGNet}}: A compact convolutional neural network for {{EEG-based}} brain--computer interfaces,''
\newblock {\em J. Neural Eng.}, vol. 15, no. 5, pp. 056013, July 2018.

\bibitem{shiConvolutionalLSTMNetwork2015}
Xingjian SHI et~al.,
\newblock ``Convolutional {{LSTM Network}}: {{A Machine Learning Approach}} for {{Precipitation Nowcasting}},''
\newblock in {\em Advances in {{Neural Information Processing Systems}}}. 2015, vol.~28, Curran Associates, Inc.

\bibitem{joshiSingleTrialP3002018}
Raviraj Joshi et~al.,
\newblock ``Single {{Trial P300 Classification Using Convolutional LSTM}} and {{Deep Learning Ensembles Method}},''
\newblock in {\em Intelligent {{Human Computer Interaction}}}, Uma~Shanker Tiwary, Ed., Cham, 2018, Lecture {{Notes}} in {{Computer Science}}, pp. 3--15, Springer International Publishing.

\bibitem{tuleuovDeepLearningModels2019}
Adilet Tuleuov and Berdakh Abibullaev,
\newblock ``Deep {{Learning Models}} for {{Subject-Independent ERP-based Brain-Computer Interfaces}},''
\newblock in {\em 2019 9th {{International IEEE}}/{{EMBS Conference}} on {{Neural Engineering}} ({{NER}})}, Mar. 2019, pp. 945--948.

\bibitem{leoniSingletrialStimuliClassification2022}
Jessica Leoni et~al.,
\newblock ``Single-trial stimuli classification from detected {{P300}} for augmented {{Brain}}--{{Computer Interface}}: {{A}} deep learning approach,''
\newblock {\em Machine Learning with Applications}, vol. 9, pp. 100393, Sept. 2022.

\bibitem{depuydtSingletrialERPQuantification2023}
Emma Depuydt et~al.,
\newblock ``Single-trial {{ERP Quantification Using Neural Networks}},''
\newblock {\em Brain Topogr}, vol. 36, no. 6, pp. 767--790, Nov. 2023.

\bibitem{polichUpdatingP300Integrative2007}
John Polich,
\newblock ``Updating {{P300}}: An integrative theory of {{P3a}} and {{P3b}},''
\newblock {\em Clin Neurophysiol}, vol. 118, no. 10, pp. 2128--2148, Oct. 2007.

\bibitem{marisNonparametricStatisticalTesting2007a}
Eric Maris and Robert Oostenveld,
\newblock ``Nonparametric statistical testing of {{EEG-}} and {{MEG-data}},''
\newblock {\em J Neurosci Methods}, vol. 164, no. 1, pp. 177--190, Aug. 2007.

\bibitem{geirnaertElectroencephalographyBasedAuditoryAttention2021}
Simon Geirnaert et~al.,
\newblock ``Electroencephalography-{{Based Auditory Attention Decoding}}: {{Toward Neurosteered Hearing Devices}},''
\newblock {\em IEEE Signal Process. Mag.}, vol. 38, no. 4, pp. 89--102, July 2021.

\end{thebibliography}

\end{document}